\documentclass[journal]{IEEEtran}
\ifCLASSINFOpdf

\else

\fi

\usepackage[cmex10]{amsmath}
\usepackage{amssymb}
\usepackage{cite}
\usepackage{graphicx}
\usepackage{array,color}
\usepackage{amsmath}
\usepackage{stfloats}
\usepackage{graphicx}
\usepackage{subfigure}
\usepackage[justification=centering]{caption}
\usepackage{multirow}
\usepackage{tabularx}
\usepackage{epsfig,epsf,color,balance,cite}
\usepackage{algorithmic}
\usepackage{algorithm}
\usepackage{bm}
\usepackage{textcomp}

\usepackage{xcolor}
\usepackage{xpatch}

\makeatletter
\def\changeBibColor#1{%
	\in@{#1}{9167249, 9140329, di2019smart, huang2020holographic, 9148961}	
	\ifin@\color{black}\else\normalcolor\fi
}
\xpatchcmd\@bibitem
{\item}
{\changeBibColor{#1}\item}
{}{\fail}

\xpatchcmd\@lbibitem
{\item}
{\changeBibColor{#2}\item}
{}{\fail}
\makeatother

\begin{document}

\title{Joint Trajectory and Passive Beamforming Design for Intelligent Reflecting Surface-Aided UAV Communications: A Deep Reinforcement Learning Approach}
\author{Liang Wang, Kezhi Wang, Cunhua Pan, and Nauman Aslam
	
	\thanks{
		Corresponding author: Kezhi Wang
		
		Liang Wang is with the School of Aerospace, Transport and Manufacturing, Cranfield University, Milton Keynes, MK43 0AL, U.K., email: liang.wang.133@cranfield.ac.uk.
		
	Kezhi Wang and Nauman Aslam are with the Department
	of Computer and Information Science, Northumbria University, Newcastle upon Tyne, NE1 8ST, U.K., emails: \{kezhi.wang, nauman.aslam\}@northumbria.ac.uk.
	
	Cunhua Pan is with the National Mobile Communications Research Laboratory, Southeast University, China., email: cpan@seu.edu.cn.

}
	
}

\maketitle
\vspace{-1.9cm}
\begin{abstract}
\textcolor{black}{In this paper, the intelligent reflecting surface (IRS)-aided unmanned aerial vehicle (UAV) communication system is studied, where the UAV is deployed to serve the user equipment (UE) with the assistance of multiple IRSs mounted on several buildings to enhance the communication quality between UAV and UE. We aim to maximize the energy efficiency of the system, including the data rate of UE and the energy consumption of UAV via jointly optimizing the UAV's trajectory and the phase shifts of reflecting elements of IRS, when the UE moves and the selection of IRSs is considered for the energy saving purpose. Since the system is complex and the environment is dynamic, it is challenging to derive low-complexity algorithms by using conventional optimization methods. To address this issue, we first propose a deep Q-network (DQN)-based algorithm by discretizing the trajectory, which has the advantage of training time. Furthermore, we propose a deep deterministic policy gradient (DDPG)-based algorithm to tackle the case with continuous trajectory for achieving better performance. The experimental results show that the proposed algorithms achieve considerable performance compared to other traditional solutions.}
\end{abstract}

\begin{IEEEkeywords}
	Deep Reinforcement Learning, UAV communications, Intelligent Reflecting Surface.
\end{IEEEkeywords}

\IEEEpeerreviewmaketitle

\section{Introduction}

It is widely envisioned that the fifth-generation (5G) wireless networks and beyond will achieve 1000-fold increase in network capacity, accommodate about 100 billion devices and support a number of emerging applications such as virtual reality (VR) services. To satisfy this ever-increasing demand, unmanned aerial vehicle (UAV) has been applied and regarded as one of the most promising technologies to achieve these ambitious goals. Compared to the traditional communication systems that utilize the terrestrial fixed base stations, UAV-aided communication systems are more cost-effective and likely to achieve better quality of service (QoS) due to its appealing properties of flexible deployment, fully controllable mobility and low cost. In fact, with the assistance of UAVs, the system performance (e.g., data rate and latency) can be significantly enhanced by establishing the line-of-sight (LoS) communication links between UAVs and user equipments (UEs). 

In addition, to further improve the channel quality, adaptive communications can be designed through the mobility/deployment control of the UAV systems. 
For example, in \cite{9275621}, Jiang \textit{et al.} proposed
a heterogeneous mobile edge computing (MEC) framework,
where ground stations (GSs), ground vehicles (GVs) and UAVs
are deployed for providing computing, communication and
caching (3C) resources at the network edge. In \cite{9167249}, Yang \textit{et al.}
investigated the weighted-sum cost minimization problem in
a hierarchical machine learning (ML) tasks distribution (HMTD) framework and
they optimized the offloading strategy, including the binary
offloading and partial offloading between the UAV and target. In~\cite{al2014optimal}, Hourani \textit{et al.} proposed an analytical approach for optimizing the altitude of UAV for the purpose of maximizing the radio coverage on the ground. In~\cite{wu2018common}, the authors considered the scenario of UAVs in an orthogonal frequency division multiple access (OFDMA) system and they proposed an iterative block coordinate descent approach for optimizing the UAV's trajectory and resource allocation, aiming to maximize the minimum average throughput of UEs. The optimization problem of UAV placement and transmit power in UAV-aided relay systems was studied in~\cite{9088229}, where Ren~\textit{et al.} proposed a low-complexity iterative algorithm to solve the problem both in the free-space channel and three-dimensional channel scenarios. In~\cite{zeng2019energy}, to minimize the energy consumption of UAV, Zeng~\textit{et al.} formulated a travelling sale problem and proposed an efficient algorithm to optimize the UAV's trajectory, including the hovering locations and duration. In~\cite{8432464}, a multi-UAV-assisted communication system was studied. The authors proposed a energy-efficient distributed MCS (Edics) algorithm to optimize the UAVs' trajectory for maximizing the energy efficiency of UAVs. \textcolor{black}{In~\cite{lu2020uav}, Lu~\textit{et al.} studied the jamming problem in UAV-aided cellular system, where the relay power is optimized without the knowledge of the cellular topology through a deep reinforcement learning (DRL) approach.} Other contributions of UAVs include their applications in MEC~\cite{8764580, 9209079, 9354996}, device-to-device communication~\cite{8514812}, data collection~\cite{8119562}, mobile crowd sensing~\cite{8664596} and wireless power transfer networks~\cite{8365881}. In~\cite{8764580}, Yang~\textit{et al.} studied the power minimization problem in a multi-UAV-enabled MEC system, where the user association, power control, computation capacity allocation and location planning were optimized. In~\cite{9209079}, Wang~\textit{et al.} investigated energy minimization problem in the multi-UAV assisted MEC system, where they proposed a multi-agent deep reinforcement learning approach for optimizing the trajectories of UAVs. In~\cite{9354996}, the authors proposed a convex optimization based trajectory (CAT) and deep Reinforcement learning based trajectory (RAT) algorithms for optimizing the user association, resource allocation and the trajectory of UAVs, aiming at minimization the energy consumption of UEs. In~\cite{8514812}, Huang~\textit{et al.} investigated the device-to-device (D2D) rate maximization problem in UAV-aided wireless communication systems, where they proposed an iterative algorithm for optimizing the UAV flying altitude, location and the bandwidth allocation, which proved that the altitude of the UAV is vital for improving the system performance. In~\cite{8664596}, Liu~\textit{et al.} introduced a distributed mobile crowed sensing platform, where multiple UAVs are deployed as mobile terminals for collecting data. They proposed a DRL-based approach for navigating a group of UAVs in order to maximize the collected data, the geographical fairness, and the energy efficiency of UAVs. In~\cite{8365881}, Xu~\textit{et al.} studied the problem of maximizing the energy harvested at all energy receivers in a UAV-enabled wireless power transfer system, in which they first proposed an algorithm based on Lagrange dual method for optimizing UAV's trajectory in an ideal case. Then, they proposed a new successive hover-and-fly algorithm based on convex programming optimization for trajectory design for the general case.  

However, in the crowded area, the communication signals between UAV and UE may be blocked by high buildings or other constructions. Thanks to the development of meta-materials or meta-surfaces~\cite{cui2014coding, di2019smart}, intelligent reflecting surface (IRS), or reconfigurable intelligent surfaces (RIS) \cite{li2017electromagnetic,9140329} has been proposed and received considerable attention in both academia and industry. In general, the IRS consists of an array of low-cost and passive reflecting elements, each of which is able to reflect the incident signals by smartly adjusting the phase shift, which has the potential to improve the achievable data rate~\cite{wu2019intelligent}. Furthermore, since the reflecting elements of the IRS can be passive, the IRS is more energy-efficient than traditional relay-aided communication techniques, such as \cite{6942282}.

Due to the above advantages, the IRS has been extensively investigated in various wireless communication systems. In~\cite{huang2020holographic}, the authors investigated the Holographic Multiple Input Multiple Output Surface (HMIMOS) architecture and analyzed its opportunities and challenges in 6G wireless networks.
In~\cite{8647620}, an IRS-enhanced MISO wireless system was studied, and the authors proposed a semidefinite relaxation (SDR) based algorithm for optimizing the active and passive beamforming, aiming to maximize the overall received signal power at the user. In~\cite{yang2020intelligent}, Yang~\textit{et al.} studied a realistic IRS-enhanced OFDM system, where the frequency-selective channels were considered, and the passive array reflecting coefficients were optimized for maximizing the achievable data rate of the user. 
In order to enhance the physical layer security of IRS-aided communication systems, Yu~\textit{et al.}~\cite{9014322} jointly optimized the beamforming at the transmitter and the phase shifts of the IRS, maximizing the physical layer security data rate. 
For multicast scenarios, the authors in~\cite{zhou2020intelligent} investigated the downlink IRS-aided multigroup multicast communication system, where the IRS can be deployed to enhance the worst-case user channel condition.  In~\cite{pan2020multicell}, Pan~\textit{et al.} studied the weighted sum rate (WSR) maximization problem for an IRS-assisted multicell MIMO communication system, and the authors proposed a pair of algorithms named Majorization-Minimization (MM) and Complex Circle Manifold (CCM) for optimizing the phase shifts of the IRS. The simulation results in~\cite{pan2020multicell} showed that the IRS is very effective in mitigating the cell-edge interference. Additionally, the authors in~\cite{9110849} considered to deploy an IRS in a simultaneous wireless information and power transfer (SWIPT) system to enhance both the energy harvesting and data rate performance. In~\cite{9133107}, the IRS was shown to be beneficial in reducing the latency of the mobile edge computing system. In~\cite{9148961}, the authors studied the achievable rate problem in an IRS-aided wireless system, and they optimized the transmit beamforming and the IRS reflect beamforming through the alternating optimization (AO) based technique. 
In~\cite{zappone2020overhead}, the authors studied the resource allocation for a point-to-point IRS-aided MIMO communication system when taking into account the channel estimation and channel feedback overhead. \textcolor{black}{In~\cite{9110869}, Huang~\textit{et al.} proposed a DRL-based algorithm to optimize the design of beamforming matrix and phase shift matrix in RIS-based multi-user MISO system.}

Against the above background, we study an IRS-aided UAV system where the UAV is deployed to provide communication services to the ground UE. To enhance the channel condition between UAV and UE, which may be blocked by some obstacles such as high buildings, the IRS may be mounted on the exterior wall of the buildings. We aim to maximize the energy efficiency of UAV, including the data rate of UE and the energy consumption of UAV via jointly optimizing the UAV's trajectory, the phase shifts of the reflecting elements of IRS, while UE moves. To address this problem, firstly, we propose a deep Q-network (DQN)-based algorithm by discretizing the trajectory for the easy deployment. Then, we propose a deep deterministic policy gradient (DDPG)-based algorithm to tackle the continuous situation for better performance. The experiment verifies that the proposed algorithms achieve better performance compared to benchmark solutions.

\begin{table}[!htpb]
	\centering
	\caption{Main Notations.}\label{tab1}
	\begin{center}
		\begin{tabular}{|c|c|}
			\hline
			\textbf{Notation} & \textbf{Definition} \\ \hline
			$k, K, \mathcal{K}$ & \text{the index, the number, and the set of IRSs}  \\ \hline
			$Z^{\text{min}}, Z^{\text{max}}$ & \text{the minimal, maximal of flying altitude of UAV}  \\ \hline
			$X^{\text{max}}, Y^{\text{max}}$ & \text{side length of target area}  \\ \hline
			$t, T, \mathcal{T}$ & \text{the index, the number, and the set of TSs}  \\ \hline
			$M_r, M_c$ & \text{number of reflecting elements of IRSs}  \\ \hline
			$x^{\text{max}}, y^{\text{max}}, z^{\text{max}}$ & \text{maximal flying distances of UAV}  \\ \hline
			$a^x_t, a^y_t, a^z_t$ & \text{flying distances of UAV in TS $t$}  \\ \hline
			$[x^u_0, y^u_0, z^u_0]$ & \text{initial coordinate of UAV}  \\ \hline
			$U_r$ & \text{tip speed of the rotor blade}  \\ \hline
			$V_h$ & \text{the mean rotor induced velocity when hovering}  \\ \hline
			$d_0$ & \text{the main body drag ratio}  \\ \hline
			$\rho_a$ & \text{air density}  \\ \hline
			$z$ & \text{the rotor solidity}  \\ \hline
			$G$ & \text{rotor disc area}  \\ \hline
			$t_d$ & \text{time duration of TS}  \\ \hline
			$[x_k, y_k, z_k]$ & \text{coordinate of IRS $k$}  \\ \hline
			$[x^e_t, y^e_t]$ & \text{coordinate of UE in TS $t$}  \\ \hline
			$[x^u_t, y^u_t, z^u_t]$ & \text{coordinate of UAV in TS $t$}  \\ \hline
			$d^{\text{UI}}_{k,t}$ & \text{distance between UAV and IRS $k$ in TS $t$}  \\ \hline
			$d^{\text{IE}}_{k,t}$ & \text{distance between UE and IRS $k$ in TS $t$}  \\ \hline
			$\bm{h}^{\text{UI}}_{k,t}$ & \text{channel gain of UAV-IRS $k$ link in TS $t$}  \\ \hline
			$\mu$ & \text{path loss at reference distance $1m$}  \\ \hline
			$\alpha^{\text{IE}}$ & \text{path loss exponent}  \\ \hline
			$f, c$ & \text{carrier frequency, speed of light}  \\ \hline
			$\bm{h}^{\text{IE}}_{k,t}$ & \text{channel gain of IRS $k$ - UE link in TS $t$}  \\ \hline
			$P, \sigma^2, B$ & \text{transmission power, noise power, bandwidth}  \\ \hline
			$\bm{\Theta}_{k,t}$ & \text{phase shift matrix of IRS $k$ in TS $t$}  \\ \hline
			$R_{k,t}$ & \text{data rate of UAV-IRS $k$-UE linke in TS $t$}  \\ \hline
		\end{tabular}
	\end{center}
\end{table}

The reminder of this paper is organized as follows. In Section~\ref{related_work}, we introduce the related work and the background of DRL. In Section~\ref{system_model}, we describe the system model, including the optimization problem. In Section~\ref{proposed_algorithm}, we present the proposed DQN and DDPG-based algorithms. In Section~\ref{simulation_result}, the experimental results are shown. Finally, we conclude the paper in Section~\ref{conclusion}. The main notations used in this paper are summarized in Table.~\ref{tab1}.

\textit{Other Notations}: In this paper, $\mathbb{C}^{M\times1}$ denotes the set of $M\times1$ complex vectors. $\text{diag}(\cdot)$ denotes the diagonalization operation. $(\cdot)^T$ denotes the transpose operation. $\mathbb{E}[\cdot]$ denotes the expectation operation.

\section{Related Work and Background}\label{related_work}

\subsection{IRS-aided UAV Communications}
Most recently, the integration of IRS in UAV-aided communication systems has become a hot research topic. For example, in~\cite{li2020reconfigurable}, the authors considered a downlink transmission system, consisting of a rotary-wing UAV, a ground user and an IRS. In this work, the authors proposed a successive convex approximation (SCA) based algorithm to optimize the UAV's trajectory and passive beamforming of the IRS. 
In \cite{ma2019enhancing}, the potential of
IRS in UAV-assisted communication systems was investigated.
The authors concluded that the deployment of IRS is capable
of achieving significant performance gain in UAV-assisted
cellular networks.
Some other benefits of IRS-assisted system can also be found in the literature. Most of the existing algorithms are based on convex optimization theory, which may achieve suboptimal performance and is time-consuming due to the fact that a number of iterations are required for the convergence of the algorithm. Their complexity may increase with the number of reflecting elements.

\subsection{DRL Background}
Thanks to the advances in the field of machine learning, most of sophisticated optimization problems may be solved efficiently and in real time. As a branch of machine learning algorithms, reinforcement learning (RL) is viewed as a useful approach for tackling complicated control tasks, such as robotics and games. In~\cite{10.5555/3312046}, Sutton~\textit{et al.} proposed a widely used model-free RL algorithm named Q-learning, where some fundamental knowledge, such as agent, environment, state, action, reward and Q-value were introduced. In addition, another mechanism named Q-table was employed in Q-learning. However, as the size of Q-table is finite, Q-learning may only handle control problems in discrete state and action spaces. As an extension of Q-learning, Mnih~\textit{et al.}~\cite{mnih2015human} proposed the deep Q-network (DQN) algorithm, which combines RL and the powerful deep neural network (DNN). Additionally, two techniques named experience replay and target network were integrated. The experimental results proved that DQN is capable of achieving enhanced performance in the challenging Atari 2600 games. In DQN, the Q-table is replaced by the DNN, as DQN can handle the control problem with infinite state spaces. However, the action space of DQN is still discrete. Inspired by DQN, Silver~\textit{et al.} proposed a deep deterministic policy gradient (DDPG)~\cite{lillicrap2015continuous} algorithm based on the actor-critic~\cite{6313077} method, which is able to be applied to continuous action spaces. Although some researchers has started to apply the DRL in the IRS or IRS-assisted UAV communications, most of the work did not consider the selection of IRS and the movement of UE. In this paper, DDPG and DQN will be applied in IRS-aided UAV system, where the selection of IRS and the movement of UE will also be considered.

\section{System Model}\label{system_model}

\begin{figure}[!htpb]
	\centering
	\includegraphics[width=3.5in]{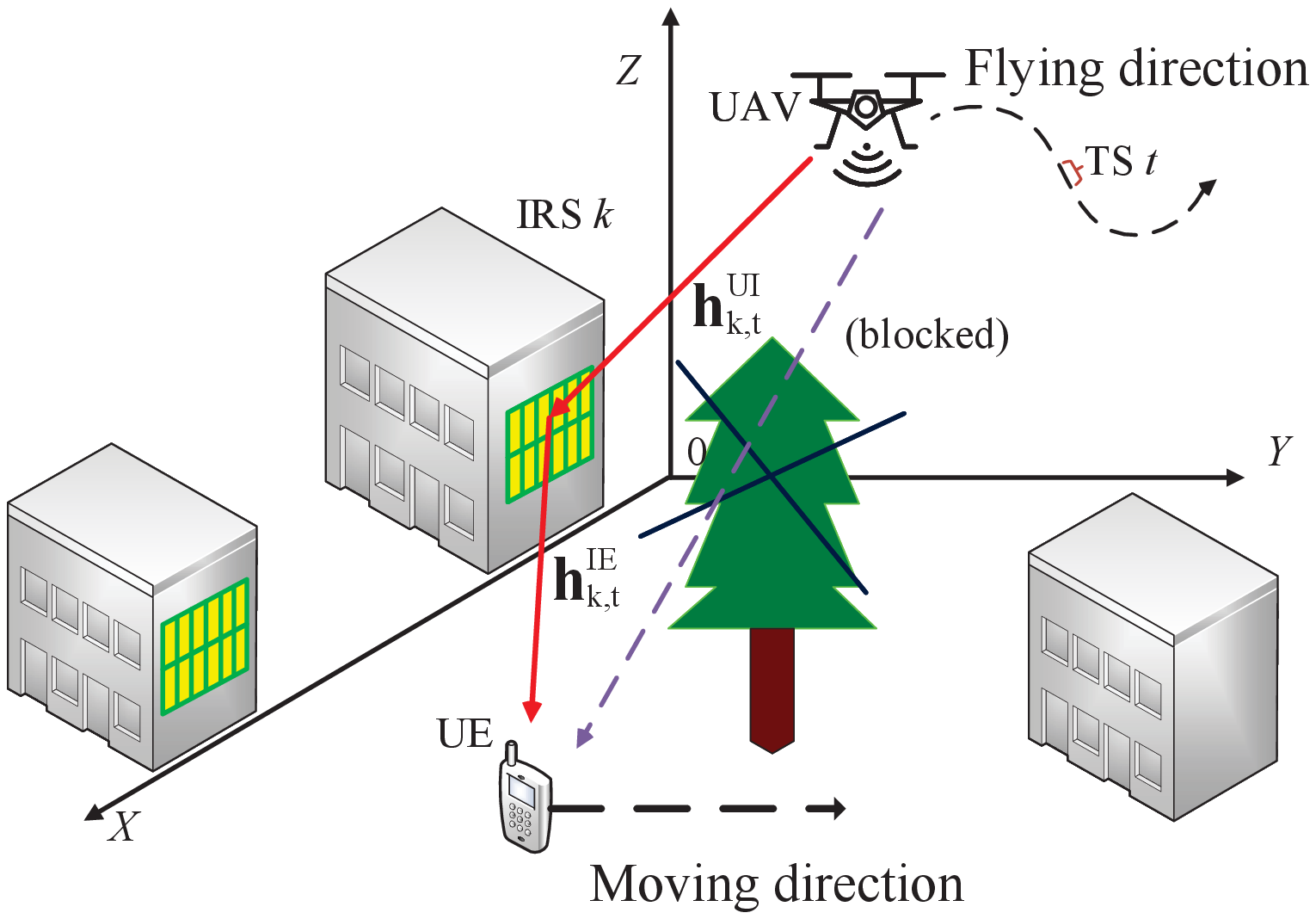}
	\caption{Architecture of IRS-aided UAV communication system}\label{dia}
\end{figure}

\textcolor{black}{Assume that there is one rotary UAV, $K$ IRSs mounted on $K$ buildings, respectively and one moving UE to be served, as shown in Fig.~\ref{dia}. Note that the UE can also be other moving object, like the autonomous vehicle.
Also, assume that the UE is located in the crowded area where it suffers from severe path loss and high attenuation, caused by high buildings and trees. Thus, the direct link between UAV and UE is not considered. IRSs are deployed for enhancing the communication quality of UE. The UAV flies within a particular altitude ranging from $[Z^\text{min}, Z^\text{max}]$ over a rectangle target area with side lengths $X^\text{max}$ and $Y^\text{max}$ for a certain number of time slots (TSs) $T$, each of which has $t_d$ time duration. For simplicity, we denote the set of IRSs as $\mathcal{K}\triangleq \{k=1,2,...,K\}$ and the set of TSs is denoted as $\mathcal{T} \triangleq \{1,2,...T\}$. Additionally, each of IRSs is equipped with an uniform rectangular array (URA) with $M_r\times M_c$ reflecting elements, which could boost the useful signal power by adjusting the phase shifts of the reflecting elements.}
\subsection{UAV model}
\textcolor{black}{In this subsection, we describe the UAV model with Cartesian coordinate system. Specifically, in each of TS, the UAV moves with a flying action determined by two horizontal distances $a^x_t \in [-x^\text{max}, x^\text{max}]$, $a^y_t \in [-y^\text{max}, y^\text{max}]$ and a vertical distance $a^z_t \in [-z^\text{max}, z^\text{max}]$. Thus, given the initial coordinate of the UAV, which is $[x^u_0, y^u_0, z^u_0]$, the coordinate of the UAV in TS $t$ is expressed as $[x^u_t, y^u_t, z^u_t]$, where $x^u_t = x^u_0 + \sum_{t'=1}^{t}a^x_{t'}$, $y^u_t = y^u_0 + \sum_{t'=1}^{t}a^y_{t'}$, and $z^u_t = z^u_0 + \sum_{t'=1}^{t}a^z_{t'}$. Note that as the UAV may not go beyond the border of the targeted area, we have the following constraints:}

\textcolor{black}{\begin{equation}
		\begin{aligned}
			0 \leq	x^u_t \leq X^\text{max},
		\end{aligned}
	\end{equation}
	and 
	\begin{equation}
		\begin{aligned}
			0 \leq	y^u_t \leq Y^\text{max},
		\end{aligned}
	\end{equation}
	and 
	\begin{equation}
		\begin{aligned}
			Z^\text{min} \leq z^u_t \leq Z^\text{max}.
		\end{aligned}
\end{equation}}

\textcolor{black}{In this work, the energy consumption for communication, such as communication circuitry and signal processing, is ignored compared with the propulsion energy. According to~\cite{zeng2019energy}, the propulsion energy consumption in TS $t$ is expressed as
	\begin{equation}\label{uav_energy}
		\begin{aligned}
			e_t = & \bigg(P_s\big(1+3(\frac{v^h_t}{U_r})^2\big)+P_m\big(\sqrt{1+\frac{1}{4}(\frac{v^h_t}{V_h})^4} - \frac{1}{2}(\frac{v^h_t}{V_h})^2 \big)^{\frac{1}{2}}\\
			& + \frac{1}{2}d_0\rho_a z G (v^h_t)^3 + P_k v^v_t\bigg)t_d,
		\end{aligned}
	\end{equation}
	where $P_s$, $P_m$ and $P_k$ are fixed constants and can be obtained from~\cite{zeng2019energy}; $U_r$ is the tip speed of the rotor blade; $V_h$ denotes the mean rotor induced velocity when hovering; $d_0$ is the main body drag ratio; $\rho_a$ is the air density; $z$ means the rotor solidity; $G$ is known as the rotor disc area; $v^h_t = \frac{\sqrt{(a^x_t)^2+(a^y_t)^2}}{t_d}$; and $v^v_t = \frac{|a^z_t|}{t_d}$.}

\subsection{Channel model}

\textcolor{black}{Denote the coordinate of IRS $k$ as $[x_k, y_k, z_k]$, the coordinate of UE as $[x^e_t, y^e_t]$. In this paper, the location of UE varies with time, as UE moves. Thus, the distance between UAV and IRS $k$ in TS $t$ is
	\begin{equation}
		\begin{aligned}
			d^{\text{UI}}_{k,t} = \sqrt{(x^u_t-x_k)^2+(y^u_t-y_k)^2+(z^u_t-z_k)^2}.
		\end{aligned}
\end{equation}}

\textcolor{black}{Similarly, the distance between IRS $k$ and UE in TS $t$ is given by
	\begin{equation}
		\begin{aligned}
			d^{\text{IE}}_{k,t} = \sqrt{(x^e_t-x_k)^2+(y^e_t-y_k)^2+(z_k)^2}.
		\end{aligned}
\end{equation}}

\textcolor{black}{Then, for the 3-D channel model, the path loss of UAV-IRS $k$ link in TS $t$ can be denoted by $P^\text{UI}_{k,t}$~\cite{al2014optimal}:
	\begin{equation}
		\begin{aligned}
			P^\text{UI}_{k,t} = \frac{A}{1+u\text{exp}(-w(\theta_{k,t}-u))}  +20\text{log}_{10}(d^\text{UI}_{k,t})+C, 
		\end{aligned}
	\end{equation}
	where $A=\eta_\text{LoS}-\eta_\text{NLoS}$, $C=20\text{log}_{10}(\frac{4\pi f}{c})+\eta_\text{NLoS}$. Note that $\eta_\text{LoS}$ and $\eta_\text{NLoS}$ are variables related to the LoS and NLoS links, respectively. $\theta_{k,t}=\text{arctan}(\frac{z^u_t-z_k}{\sqrt{(x^u_t-x_k)^2+(y^u_t-y_k)^2}})$ denotes the elevation angle between UAV and IRS $k$ in TS $t$. $f$, $c$ are the carrier frequency and speed of light, respectively. $u$ and $w$ are constant values determined by the environment. Thus, motivated by~\cite{al2014optimal},  the channel gain of UAV-IRS $k$ link in TS $t$ is denoted by $\bm{h}^{\text{UI}}_{k,t} \in \mathbb{C}^{M_r M_c \times 1}$:}
\textcolor{black}{\begin{equation}
		\begin{aligned}
			\bm{h}^{\text{UI}}_{k,t} = \widetilde{C}(d^{\text{UI}}_{k,t})^{-2}e^{\frac{\widetilde{A}}{1+a\text{exp}(-b(\theta_{k,t}-a))}}\hat{\bm{h}}^{\text{UI}}_{k,t},
		\end{aligned}
	\end{equation}
	where $\widetilde{C}=10^{-\frac{C}{10}}$, $\widetilde{A}=-A\frac{\text{ln}10}{10}$. $\hat{\bm{h}}^{\text{UI}}_{k,t}\in \mathbb{C}^{M_r M_c \times 1}$ is the LoS component~\cite{ren2019achievable, wei2020sum}:
	\begin{equation}
		\begin{aligned}
			\hat{\bm{h}}^{\text{UI}}_{k,t} =& \big[1, e^{-j\frac{2\pi}{\lambda}d\phi^{\text{UI}}_{k,t}\varphi^{\text{UI}}_{k,t}},..., e^{-j\frac{2\pi}{\lambda}(M_r-1)d\phi^{\text{UI}}_{k,t}\varphi^{\text{UI}}_{k,t}}  \big]^T\\
			&\otimes\big[1, e^{-j\frac{2\pi}{\lambda}d\psi^{\text{UI}}_{k,t}\varphi^{\text{UI}}_{k,t}},..., e^{-j\frac{2\pi}{\lambda}(M_c-1)d\psi^{\text{UI}}_{k,t}\varphi^{\text{UI}}_{k,t}}  \big]^T,
		\end{aligned}
	\end{equation}
	in which $\lambda$ is the carrier wavelength, $d$ is the antennas separation distance. $\phi^{\text{UI}}_{k,t}=\frac{x^u_t-x_k}{d^{\text{UI}}_{k,t}}$, $\psi^{\text{UI}}_{k,t} =\frac{y_k-y^u_t}{d^{UI}_{k,t}}$, $\varphi^{\text{UI}}_{k,t} = \frac{z^u_t - z_k}{d^{\text{UI}}_{k,t}}$ represent the cosine, sine values of the horizontal, vertical angles of arrival (AoA) of the signal from the UAV to IRS $k$ in TS $t$, respectively.}

\textcolor{black}{Furthermore, the channel gain of IRS $k$ - UE link in TS $t$, is denoted by $\bm{h}^{\text{IE}}_{k,t} \in \mathbb{C}^{M_r M_c \times 1}:$
	\begin{equation}
		\begin{aligned}
			\bm{h}^{\text{IE}}_{k,t}=&\sqrt{\frac{\mu}{(d^{\text{IE}}_{k,t})^{\alpha^{\text{\text{IE}}}}}}\hat{\bm{h}}^{\text{IE}}_{k,t},
		\end{aligned}
	\end{equation}
	in which $\mu$ is the path loss at the reference distance $1m$, $\alpha^{\text{\text{IE}}}$ is the path loss exponent. $\hat{\bm{h}}^{\text{IE}}_{k,t} \in \mathbb{C}^{M_r M_c \times 1}$ is the LoS component which is given by~\cite{wei2020sum}:
	\begin{equation}
		\begin{aligned}
			\hat{\bm{h}}^{\text{IE}}_{k,t}=& \big[1, e^{-j\frac{2\pi}{\lambda}d\phi^{\text{IE}}_{k,t}\varphi^{\text{IE}}_{k,t}},..., e^{-j\frac{2\pi}{\lambda}(M_r-1)d\phi^{\text{IE}}_{k,t}\varphi^{\text{IE}}_{k,t}}  \big]^T\\
			&\otimes\big[1, e^{-j\frac{2\pi}{\lambda}d\psi^{\text{IE}}_{k,t}\varphi^{\text{IE}}_{k,t}},..., e^{-j\frac{2\pi}{\lambda}(M_c-1)d\psi^{\text{IE}}_{k,t}\varphi^{\text{IE}}_{k,t}}  \big]^T,
		\end{aligned}
	\end{equation}
	where $\phi^{\text{IE}}_{k,t}=\frac{x^e_t-x_k}{d^{\text{IE}}_{k,t}}$, $\psi^{\text{IE}}_{k,t}=\frac{y^e_t-y_k}{d^{\text{IE}}_{k,t}}$, $\varphi^{\text{IE}}_{k,t}=\frac{z_k}{d^{\text{IE}}_{k,t}}$ represent the cosine, sine values of the horizontal, vertical angles of departure (AoD) of the signal from IRS $k$ to UE in TS $t$, respectively.}
\textcolor{black}{Similar to~\cite{wei2020sum}, we denote each of IRS has $M_r \times M_c$ reflecting elements, each of which can passively adjust its phase shift $\theta_{k,m_r,m_c,t} \in [-\pi, \pi)$. Thus, the diagonal phase shift matrix of IRS $k$ in TS $t$ can be expressed as $\bm{\Theta}_{k,t} = \text{diag}\big(e^{j\theta_{k,1,1,t}},...,e^{j\theta_{k,m_r,m_c,t}},...,e^{j\theta_{k,M_r,M_c,t}}\big)\in \mathbb{C}^{M_rM_c\times M_rM_c}$.}
\textcolor{black}{Then, the achievable data rate of UAV - IRS $k$ - UE link in TS $t$ is
	\begin{equation}
		\begin{aligned}
			R_{k,t} = B \text{log}_2(1+\frac{P(\bm{h}^{\text{IE}}_{k,t})^T\bm{\Theta}_{k,t}\bm{h}^{\text{UI}}_{k,t}}{B\sigma^2}),
		\end{aligned}
	\end{equation}
	where $P$ and $\sigma^2$ are the transmission and noise power respectively. $B$ is the bandwidth.}

\textcolor{black}{In this paper, assume that the UE is served with a time-division-multiple-access (TDMA) mode, where only one IRS is selected in each TS. This is very useful to save the energy consumption of the IRSs, as only one IRS may switch on at each time, whereas other IRSs may be switched off or in the sleep mode. We denote $c_{k,t}=\{0, 1\}$ as the schedule factor between UE and IRS $k$ in TS $t$, where $c_{k,t}=1$ means IRS $k$ is selected by the UE and otherwise $c_{k,t}=0$. Then, the schedule scheme is described as follows:
	\begin{equation}\label{select_irs}
		\begin{aligned}
			c_{k,t}=
			\begin{cases}
				1, k= \text{argmin}(\bm{d}^{\text{IE}}_t), \\
				0, \text{otherwise},
			\end{cases}
		\end{aligned}
	\end{equation}
	where $\bm{d}^{\text{IE}}_t=\{d^{\text{IE}}_{k,t}, \forall k \in \mathcal{K}\}$ denotes the set of distances between UE and IRSs in TS $t$. Additionally, one may have
	\begin{equation} \label{select_ir1s}
		\sum_{k=1}^{K} c_{k,t} = 1, \forall t \in \mathcal{T}.
\end{equation}}
which means that only one IRS is selected at each time. 
 Note that other selection schemes may also be applied. For example, one may select IRS based on the cascaded channel between UAV and UE. If the selection scheme is determined, i.e., (\ref{select_irs}), then (\ref{select_ir1s}) may not be needed.
\subsection{Problem Formulation}
\textcolor{black}{In this paper, we aim to maximize the energy efficiency of UAV, including the data rate of UE and energy consumption of UAV, which can be formulated as the following optimization problem.
	\begin{subequations}\label{optpro}
		\begin{IEEEeqnarray}{s,lCl'lCl'lCl}
			& \IEEEeqnarraymulticol{9}{l}{\mathcal{P}:\underset{\bm{\Theta},\bm{Z}}{\text{max}} \sum_{t=1}^{T}\frac{\sum_{k=1}^{K}c_{k,t}R_{k,t}}{e_t},
			} \\
			& \text{subject to:} \nonumber\\
			& -x^{\text{max}} \leq a^x_t \leq x^{\text{max}},\\
			& -y^{\text{max}} \leq a^y_t \leq y^{\text{max}},\\
			& -z^{\text{max}} \leq a^z_t \leq z^{\text{max}},\\
			& 0 \leq x^u_t \leq X^{\text{max}},\\
			& 0 \leq y^u_t \leq Y^{\text{max}},\\
			& Z^{\text{min}} \leq z^u_t \leq Z^{\text{max}}, \\
			& -\pi \leq \theta_{k,m_r,m_c,t} < \pi,\\
			& c_{k,t}=
			\begin{cases}
				1, k= \text{argmin}(\bm{d}^{\text{IE}}_t), \\
				0, \text{otherwise},
			\end{cases}
		\end{IEEEeqnarray}
	\end{subequations}
	where $\bm{\Theta} = \{\bm{\Theta}_{k,t},~\forall k \in \mathcal{K}, t \in \mathcal{T}\}$ and $\bm{Z} = \{[a^x_t, a^y_t, a^z_t],~\forall t \in \mathcal{T}\}$. It is quite difficult to solve the above problem in general since it is non-convex. Thus, we first propose a DQN-based algorithm to tackle the trajectory of UAV by discretizing the variables $\bm{Z}$. This has advantages in terms of training time, although it may result in a little bit of performance loss. We also propose a DDPG-based algorithm to optimize $\bm{Z}$ with continuous actions for better performance. We also show a low-complexity phase alignment scheme to optimize $\bm{\Theta}$.}

\section{Proposed algorithms}\label{proposed_algorithm}

\subsection{DQN-based Algorithm for Discrete Cases}\label{dqn_solution}

\textcolor{black}{In this subsection, we show the DQN-based algorithm. We first introduce the state, action and reward. Then, we model the whole IRS-aided UAV communication system as an environment. It is assumed that the agent is employed for interacting with the environment for the purpose of finding the optimal actions that can maximize the accumulated rewards $R_t=\sum_{t'=t}^{T}\gamma^{t'-t}r_{t'}$ within a sequence of states, where $\gamma \in [0,1]$ is the discount factor. We define the state $s_t$, the action $a_t$, and the reward $r_t$ in TS $t$ as follows.
	\begin{enumerate}
		\item State $s_t$: the state of agent in TS $t$ has the following components:
		\begin{enumerate}
			\item the coordinate of UAV: $[x^u_t, y^u_t, z^u_t]$.
			\item UAV's remaining energy level: $e^\text{max}-\sum_{t'=1}^{t}e_{t'}$, where $e^{\text{max}}$ is the maximal energy level of UAV.
			\item the index of TS: $t$.
			\item the coordinate of UE: $[x^e_t, y^e_t]$.
			\item the set of IRSs' coordinates: $\{x_k, x_k, z_k, \forall k \in \mathcal{K}\}$.
		\end{enumerate}
		\item Action $a_t$: we define the flying distances of UAV in TS $t$ as action, which is $a_t = [a^x_t, a^y_t, a^z_t]$.
		\item Reward $r_t$: we define the reward function as follows:
		\begin{equation}\label{reward}
			\begin{aligned}
				r_t = \frac{\sum_{k=1}^{K}c_{k,t}R_{k,t}}{e_t} - p,
			\end{aligned}
		\end{equation}
		where $p$ is defined as the penalty if the UAV flies out of the target area, i.e., (\ref{optpro}e), (\ref{optpro}f) or (\ref{optpro}g) are not satisfied.
\end{enumerate}}

\begin{figure}[!htpb]
	\centering
	\includegraphics[width=3.5in]{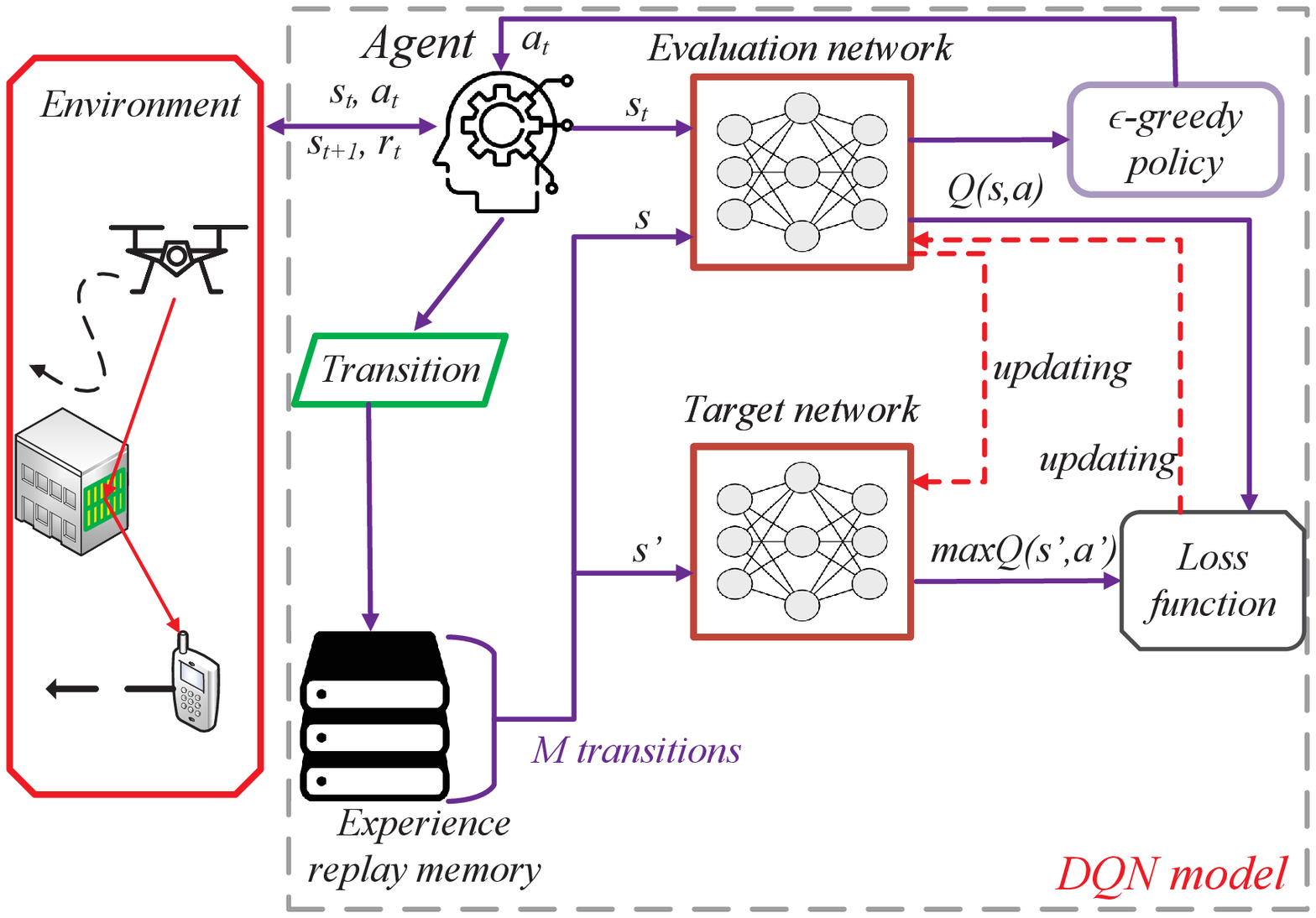}
	\caption{Architecture of DQN algorithm}\label{dqndia}
\end{figure}

\textcolor{black}{Motivated by the work that is done in~\cite{mnih2015human}, here we propose the DQN-based algorithm for optimizing the UAV's trajectory, whose overall architecture is shown in Fig.~\ref{dqndia}. In DQN, there is an agent which controls the UAV for interacting with the environment. We assume there are two DNNs named the evaluation network and target network. Note that the target network has the same structure as the evaluation network but it only updates periodically. Firstly, the agent receives the state $s_t$ from the environment and sends it to the evaluation network, which generates the Q-values $Q(s,a)$ of all actions. Based on the Q-values and following an $\epsilon$-greedy policy, the action $a_t$ is generated. After that, the reward $r_t$ is obtained from the environment. It is worth mentioning that the proposed DQN-based algorithm can only optimize the UAV's trajectory in the finitely discrete action space. Hence, we define the action space in each of TS as $\mathcal{A}$, which has the following actions:
	\begin{equation}
		\begin{aligned}
			\mathcal{A} = \begin{cases}
				[x^{\text{max}}, 0, 0], \\
				[-x^{\text{max}}, 0, 0], \\
				[\frac{x^{\text{max}}}{2}, 0, 0], \\
				[-\frac{x^{\text{max}}}{2}, 0, 0], \\
				[0, y^{\text{max}}, 0], \\
				[0, -y^{\text{max}}, 0], \\
				[0, \frac{y^{\text{max}}}{2}, 0], \\
				[0, -\frac{y^{\text{max}}}{2}, 0], \\
				[0, 0, z^{\text{max}}], \\
				[0, 0, -z^{\text{max}}], \\
				[0, 0, 0].
			\end{cases}
		\end{aligned}
\end{equation}}

\textcolor{black}{Then, the transition, which consists of $\{s_t,a_t,r_t,s_{t+1}\}$ is stored into an experience replay memory. When the experience replay memory has enough transitions, the learning procedure starts. A mini-batch randomly samples $M$ transitions to train the evaluation network. Precisely, given the Q-values $Q(s,a)$ from the evaluation network and the maximal Q-values $\text{max}Q(s',a')$ from the target network, the loss function can be calculated for updating the evaluation network, which can be expressed as
	\begin{equation}\label{dqnloss}
		\begin{aligned}
			L_i(\delta_i) = \mathbb{E}_{s,a}\bigg[\bigg(r+\gamma \underset{a'}{\text{max}}Q(s',a'|\delta_{i-1})-Q(s,a|\delta_i)\bigg)^2\bigg],
		\end{aligned}
	\end{equation}
	where $\delta$ is the parameter of DNN, and $i$ is the index of iteration.}

\textcolor{black}{\begin{algorithm}[!h]
		\caption{DQN-based algorithm}\label{algorithmDQN}
		\begin{algorithmic}[1]	
			\STATE Initialize evaluation, target networks with parameters $\delta$; \
			\STATE Initialize experience replay memory;\
			\FOR{Episode = 1,2,...,$N^{\text{eps}}$}
			\STATE Initialize state $s_t$;\
			\FOR{TS $t = 1,2,...T$}
			\STATE Obtain $s_t$;\
			\STATE Select $a_t = \underset{a_t\in\mathcal{A}}{\text{argmax}Q(s_t,a_t)}$ with probability $\epsilon$;\
			\STATE Randomly select $a_t$ from $\mathcal{A}$ with probability $1-\epsilon$;\
			\STATE Execute $a_t$;\
			\STATE Calculate the energy consumption of UAV $e_t$ from Eq.~(\ref{uav_energy});\
			\STATE Obtain the optimized phase shifts of selected IRS $k$ according to Section~\ref{ph_opt};\
			\STATE Calculate $r_t$ according to Eq.~(\ref{reward});\
			\STATE Store transition $\{s_t,a_t,r_t,s_{t+1}\}$ into experience replay memory;\
			\IF{the learning process starts}
			\STATE Randomly sample $M$ transitions from experience replay memory;\
			\STATE Update evaluation network from Eq.~(\ref{dqnloss});\
			\STATE Update target network periodically;\
			\ENDIF
			\ENDFOR
			\ENDFOR
		\end{algorithmic} 
\end{algorithm}}

\textcolor{black}{In Algorithm~\ref{algorithmDQN}, we provide the overall pseudo code of the proposed DQN algorithm. From Line 1 to 2, we initialize the evaluation, target networks and the experience replay memory. During each episode, we first initialize the state $s_t$. Then, in each TS, the agent follows an $\epsilon$-greedy policy to generate $a_t$. Precisely, the agent selects $a_t$ that has the maximal Q-value with probability $\epsilon$, or randomly selects $a_t$ from $\mathcal{A}$ with probability $1-\epsilon$. The energy consumption of UAV is calculated by Eq.~(\ref{uav_energy}). Note that in Line 11, the selection of IRS is based on Eq.~(\ref{select_irs}), and the optimization of phase shifts is introduced in Section~\ref{ph_opt}. Then, the reward is obtained by Eq.~(\ref{reward}). In Line 13, the transition will be stored into experience replay memory. From Line 14, the learning process starts with randomly sampling $M$ transitions from memory for training the evaluation network, whose parameter is updated by Eq.~(\ref{dqnloss}). Finally, the target network is also updated periodically.}

\subsection{DDPG-based Algorithm For Continuous Cases}\label{ddpg_solution}

\begin{figure}[!htpb]
	\centering
	\includegraphics[width=3.5in]{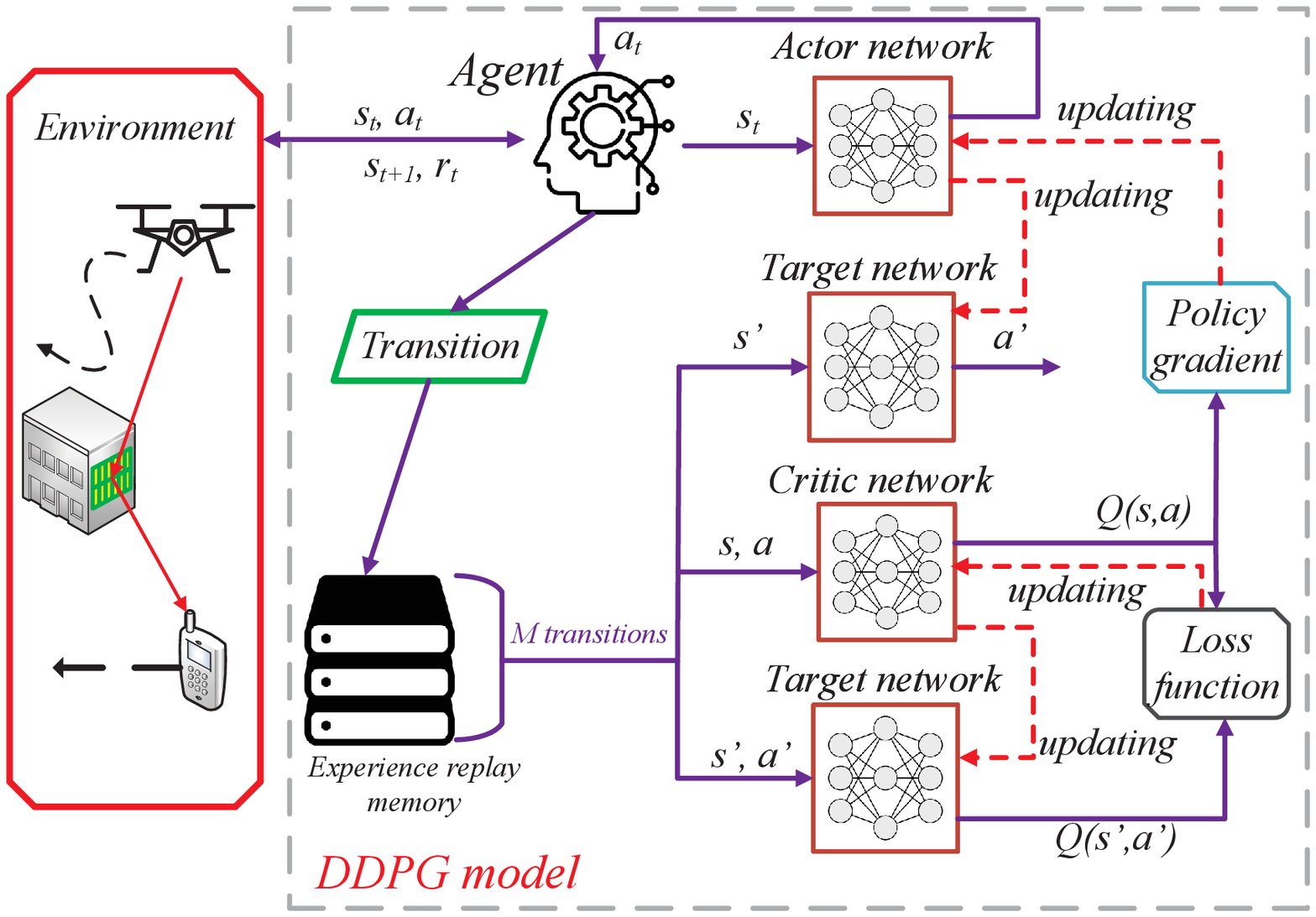}
	\caption{Architecture of DDPG algorithm}\label{ddpgdia}
\end{figure}

\textcolor{black}{In this subsection, we show the DDPG-based algorithm for tackling the continuous case and optimizing the UAV's trajectory, which applies the well-known actor-critic approach. We also show the architecture of DDPG algorithm in Fig.~\ref{ddpgdia}. There are two DNNs named actor network with function $a=\pi(s|\delta^{\pi})$ and critic network with function $Q(s,a|\delta^{Q})$ respectively. Note that $\pi(\cdot)$ maps the state and action, $Q(\cdot)$ is the approximator for generating Q-value with the given the state-action pairs. Also, there are two target networks with function $\pi'(\cdot)$ and $Q'(\cdot)$, which have the same structure with actor and critic networks, respectively. The agent receives the state $s_t$ from the environment and sends the action $a_t$ generated by its actor network. Then, the transition is stored into the experience replay memory. When the learning process starts, $M$ transitions are sampled to train the actor and critic networks. Precisely, given the states $s$ and actions $a$, the critic network generates the Q-values $Q(s,a)$ for calculating the policy gradient~\cite{lillicrap2015continuous}, which is expressed as:
	\begin{equation}\label{ddpgpolicy}
		\begin{aligned}
			\nabla_{\delta^{\pi}} J =& \mathbb{E}\big[\nabla_a Q(s,a|\delta^Q)|_{s=s_t,a=\pi(s_t|\delta^{\pi})}\\& \cdot \nabla_{\delta^{\pi}}\pi(s|\delta^{\pi})|_{s=s_t}\big].
		\end{aligned}
\end{equation}}

\textcolor{black}{Once the policy gradient is calculated, the parameter of actor network is enabled to be updated. Furthermore, the critic network is trained by the loss function~\cite{lillicrap2015continuous} as
	\begin{equation}\label{ddpgloss}
		\begin{aligned}
			L(\delta^Q) = \frac{1}{M}\sum_{m=1}^{M}\big(y_m-Q(s_m,\pi(s_m|\delta^{\pi})|\delta^{Q})^2\big),
		\end{aligned}
	\end{equation}
	where $m$ is the index of transitions in mini-batch, and $y_m = r_m+\gamma Q'(s'_m,\pi'(s'_m|\delta^{\pi'})|\delta^{Q'})$.}

\textcolor{black}{\begin{algorithm}[!h]
		\caption{DDPG-based algorithm}\label{algorithmDDPG}
		\begin{algorithmic}[1]	
			\STATE Initialize actor $\pi(\cdot)$ and critic $Q(\cdot)$ network with parameters $\delta^{\pi}$ and $\delta^{Q}$ respectively; \
			\STATE Initialize target networks $\pi'(\cdot)$, $Q'(\cdot)$ with parameters $\delta^{\pi'}=\delta^{\pi}$, $\delta^{Q'}=\delta^{Q}$;\
			\STATE Initialize experience replay memory;\
			\FOR{Episode = 1,2,...,$N^{\text{eps}}$}
			\STATE Initialize state $s_t$;\
			\FOR{TS $t = 1,2,...,T$}
			\STATE Obtain $s_t$;\
			\STATE Select $a_t = \pi(s_t|\delta^{\pi}) + \omega N'$;\
			\STATE Execute $a_t$;\
			\STATE Calculate the energy consumption of UAV $e_t$ from Eq.~(\ref{uav_energy});\
			\STATE Obtain the optimized phase shifts of selected IRS $k$ according to Section \ref{ph_opt};\
			\STATE Calculated $r_t$ according to Eq.~(\ref{reward});\
			\STATE Store transition $[s_t,a_t,r_t,s_{t+1}]$ into experience replay memory;\
			\IF{the learning process starts}
			\STATE Randomly sample $M$ transitions from experience replay memory;\
			\STATE Update critic network according to Eq.~(\ref{ddpgloss});\
			\STATE Update actor network according to Eq.~(\ref{ddpgpolicy});\
			\STATE Update two target networks with rate of $\tau$;\
			\ENDIF
			\ENDFOR
			\ENDFOR
		\end{algorithmic} 
\end{algorithm}}

We further provide the pseudo code of the proposed algorithm in Algorithm~\ref{algorithmDDPG}. From Line 1 to 3, we first initialize actor and critic networks with parameters $\delta^{\pi}$ and $\delta^Q$ respectively. Besides, two target networks and the experience replay memory are initialized as well. During each training episode, the state $s_t$ is initialized in the first TS. Then, the agent obtains $s_t$ from environment and receives $a_t$ generated by the actor network. Note that in Line 8, a random action noise $N'$ is deployed and it decays with rate of $\omega$ for better exploration. \textcolor{black}{In this paper, as the activation function of output layer of actor network is $\textit{tanh}(\cdot)$, the action $a_t$ can be expressed by $a_t = [o^x_tx^{\text{max}}, o^y_ty^{\text{max}}, o^z_tz^{\text{max}}]$, where $o^x_t, o^y_t, o^z_t$ are the output values of actor network.} After executing the action $a_t$, the consumed energy $e_t$ of UAV is obtained from Eq.~(\ref{uav_energy}). In Line 11, the optimized phase shifts of selected IRS and reward $r_t$ are obtained by Section~\ref{ph_opt} and Eq.~(\ref{reward}) respectively. After that, the transition $\{s_t,a_t,r_t,s_{t+1}\}$ is stored into the experience replay memory. When the learning process starts, the mini-batch randomly samples $M$ transitions to train the actor and critic network by Eq.~(\ref{ddpgpolicy}) and Eq.~(\ref{ddpgloss}) respectively. Additionally, two target networks are updated with the rate of $\tau=0.001$.

\subsection{Phase Shift Optimization}\label{ph_opt}

\textcolor{black}{Here, we show a low-complexity algorithm for optimizing the phase shifts of selected IRS. Specifically, by given the coordinates of UAV and UE in TS $t$, the phase shift of reflecting element $\theta_{k,m_r,m_c,t}$ of selected IRS $k$ in TS $t$ requires to be aligned, for maximizing the data rate of UE. According to~\cite{wei2020sum}, the optimal phase shift $\theta_{k,m_r,m_c,t}$ can be calculated by
	\begin{equation}
		\begin{aligned}
			\theta_{k,m_r,m_c,t} =& \frac{2\pi}{\lambda}\{d(m_r-1)\phi^{\text{IE}}_{k,t}\varphi^{\text{IE}}_{k,t}+d(m_c-1)\psi^{\text{IE}}_{k,t}\varphi^{\text{IE}}_{k,t} \\& + d(m_r-1)\phi^{\text{UI}}_{k,t}\varphi^{\text{UI}}_{k,t} +  d(m_c-1)\psi^{\text{UI}}_{k,t}\varphi^{\text{UI}}_{k,t}\}.
		\end{aligned}
\end{equation}}

\section{Simulation Result}\label{simulation_result}

\textcolor{black}{In this section, extensive simulations are conducted to evaluate the performance of the proposed algorithms. The simulation is executed in Python 3.7 and Tensorflow 1.15.0. For DQN-based algorithm, we deploy two fully-connected hidden layers with $[256, 256]$ neurons and the AdamOptimizer is used to update the evaluation network with the rate of $0.001$. While the target network is updated with 300 iterations. For DDPG-based algorithm, we also deploy two fully-connected hidden layers with $[256, 256]$ neurons in both actor and critic networks. The AdamOptimizer is used to train the actor and critic networks with the rate of 0.001. The size of experience replay memory and mini-batch are $100000$ and $128$ respectively. The coordinates of IRSs are set as $[100,-100,50]$, $[300,400,50]$, $[500,-100,50]$. In each training episode, the UAV always starts to serve UE from the initial coordinate $[0, 300, 50]$. The number of TSs is set as 100. The UE (which can be an autonomous vehicle) moves with a fixed speed 6 $\text{m/s}$ starting from the initial coordinate $[0, 150]$ to the final coordinate $[600, 150]$. Other parameters can be found in Table.~\ref{para}.}

\begin{table}[!htpb]
	\centering
	\caption{Main Notations.}\label{para}
	\begin{tabular}{|c|c|c|c|}
		\hline
		\textbf{Notation} & \textbf{Description} &\textbf{Notation} & \textbf{Description}\\ \hline
		$K$ & \text{3} & $Z^{\text{min}}$ & \text{50 m}  \\ \hline
		$Z^{\text{max}}$ & \text{300 m} & $X^{\text{max}}$ & \text{600 m}  \\ \hline
		$Y^{\text{max}}$ & \text{300 m} & $T$ & \text{100}  \\ \hline
		$x^{\text{max}}$ & \text{40 m} & $y^{\text{max}}$ & \text{40}  \\ \hline
		$z^{\text{max}}$ & \text{10 m} & $P_s$ & \text{79.85}  \\ \hline
		$P_m$ & \text{88.63} & $P_k$ & \text{11.46}  \\ \hline
		$U_r$ & \text{120 \text{m/s}} &$V_h$ & \text{4.03} \\ \hline		
		$d_0$ & \text{0.6} &$\rho_a$ & \text{1.225 $\text{kg/m}^3$} \\ \hline
		$z$ & \text{0.05} &$G$ & \text{0.503 $\text{m}^2$}  \\ \hline
		$t_d$ & \text{1 s} &$\eta_\text{LoS}$ & \text{0.1 dB}  \\ \hline
		$\eta_\text{NLoS}$ & \text{21 dB} &$u$ & \text{12.08}  \\ \hline
		$w$ & \text{0.11} &$f$ & \text{2.5 Ghz}  \\ \hline
		$c$ & \text{3$\times$ $10^8m/s$} &$d$ & \text{$\lambda/2$} \\ \hline
		$\mu$ & \text{-30 dB} & $\alpha^{\text{IE}}$ &\text{2.5}\\ \hline
		$P$ & \text{0.001 W} & $\sigma^2$ & \text{-173 dBm/Hz}  \\ \hline
		$B$ & \text{1 KHz} & $\gamma$ & \text{0.99} \\ \hline
		$e^{\text{max}}$ & \text{20000 J} & $p$ & \text{100} \\ \hline
		$\epsilon$ & \text{0.9} & $N^{\text{eps}}$ & \text{6000} \\ \hline
		$\omega$ & \text{0.99995} & $N'$ & \text{1} \\ \hline
	\end{tabular}
\end{table}

\textcolor{black}{For comparison, we present two benchmark algorithms as follows:
	\begin{itemize}
		\item Random movement and random phase shifts (RR): In this setting, the UAV randomly selects the flying action in each TS. Also, it randomly selects the phase shift for each reflecting element.
		\item Fixed movement and fixed phase shifts (FF): In this setting, the UAV moves from the initial coordinate $[0, 300, 50]$ to the final coordinate $[600, 0, 50]$. Additionally, the phase shift of each reflecting element is fixed as $\frac{\pi}{2}$.
\end{itemize}}

\begin{figure}[!htpb]
	\centering	
	\subfigure[]{	
		\begin{minipage}[b]{0.5\textwidth}\label{dqnreward}	
			\includegraphics[width=3.5in]{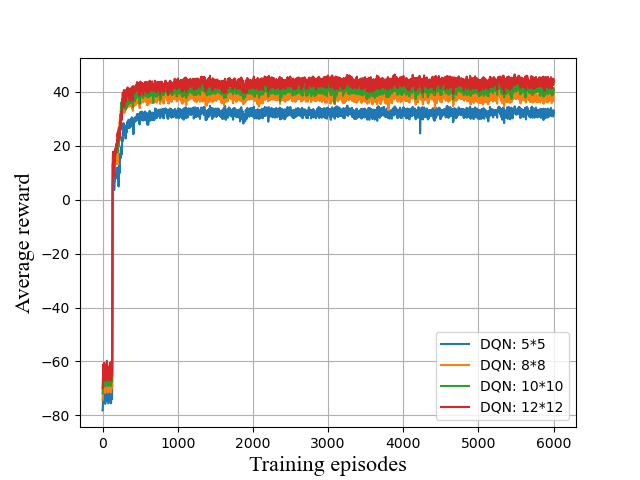}		
		\end{minipage}	
	}
	\subfigure[]{
		\begin{minipage}[b]{0.5\textwidth}\label{ddpgreward}
			\includegraphics[width=3.5in]{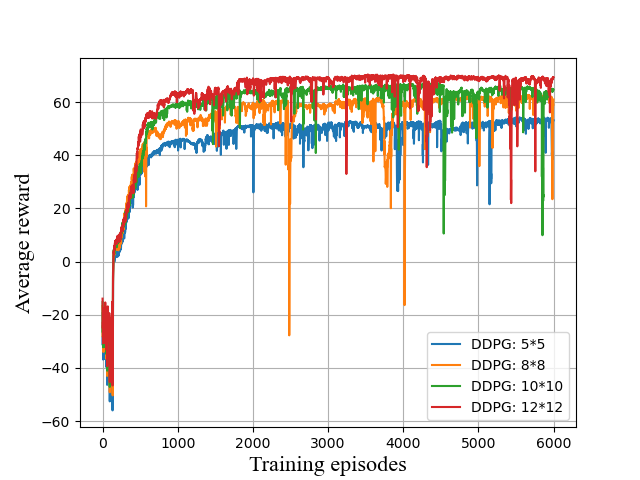}
		\end{minipage}
	}
	\caption{Average reward of (a) DQN and (b) DDPG versus the number of training episodes.} \label{rewardcomp}
\end{figure}

\textcolor{black}{First, we depict the average reward of the proposed DQN-based and DDPG-based algorithms of the training procedure with different number of reflecting elements in Fig.~\ref{rewardcomp}, where the number of IRSs is set to 3 and the number of the reflecting elements is the same for the IRSs. As shown in Fig.~\ref{dqnreward}, one can see for different number of reflection elements, the training curves of average rewards always remain negative at the beginning. This is because the UAV may have poor performance, such as flying out of the target area, resulting in negative reward. After that, as the networks start to converge, the average rewards increase and eventually remain stable, which indicate that the system find the best performance. Besides, one can observe that as the number of reflecting elements increases, the average rewards increase as well. Then, in Fig.~\ref{ddpgreward}, we depict the average rewards of the proposed DDPG-based algorithm versus the number of training episodes, which have the similar trend as DQN-based solution in Fig.~\ref{dqnreward}. It is worth noting that when the numbers of reflecting elements are the same, DDPG-based solution achieves higher reward than DQN-based solution, as expected. This is because for DQN-based algorithm, it only tries limited set of actions, whereas DDPG-based solutions optimize the variables continuously.}

\textcolor{black}{When the training is done, the networks in DQN and DDPG are saved for testing. Here, we also give the complexity of proposed DQN and DDPG-based algorithms in testing phase. Specifically, as the fully-connected layers are applied in the experiments, the complexity for networks in DQN and DDPG is $\mathcal{O}(\sum_{l=1}^{L}n_{l-1}n_l)$, where $L$ denotes the number of layers and $n_l$ is the number of neurons in $l$-th layer. Besides, the complexity for phase shift optimization in each TS is $\mathcal{O}(M_rM_c)$. Thus, the overall complexity for DQN and DDPG is $\mathcal{O}(T(M_rM_c+\sum_{l=1}^{L}n_{l-1}n_l))$. }

\begin{figure}[!htpb]
	\centering
	\includegraphics[width=3.5in]{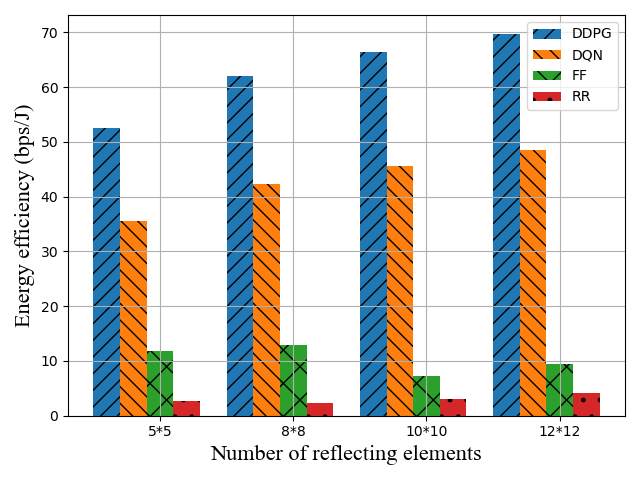}
	\caption{Average energy efficiency achieved by DQN, DDPG, FF and RR with different number of reflecting elements.}\label{ene_comp}
\end{figure}

\textcolor{black}{Then, we evaluate the performance of proposed DQN and DDPG-based algorithms. In Fig.~\ref{ene_comp}, we depict the average energy efficiency of UAV obtained by DQN, DDPG, FF, RR respectively in one episode. Specifically, the energy efficiency of UAV obtained by DDPG consistently increases from 52 $\text{bps/J}$ to 70 $\text{bps/J}$. Additionally, it is observed that for different number of reflecting elements, DDPG always achieves higher energy efficiency comparing with other algorithms. DQN performs slightly worse than DDPG, which also outperforms FF and RR.}

\begin{figure*}[!htpb]
	\centering
	\subfigure[DQN: 3D]{
		\includegraphics[width=0.49\textwidth]{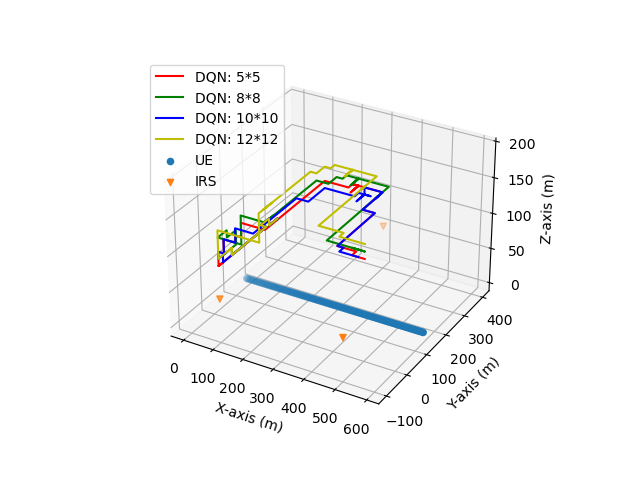}\label{dqn_3d}
	}
	\subfigure[DQN: 2D]{\includegraphics[width=0.49\textwidth]{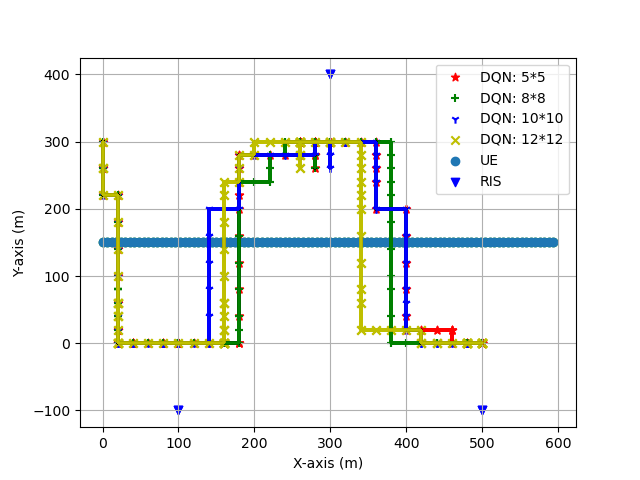}\label{dqn_2d}}
	\subfigure[DDPG: 3D]{\includegraphics[width=0.49\textwidth]{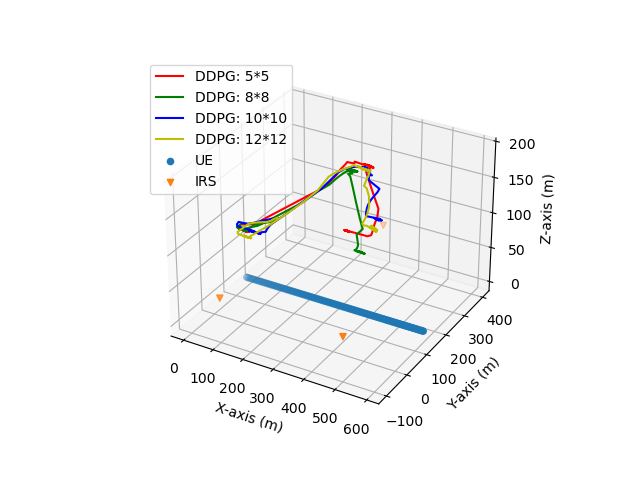}\label{ddpg_3d}}
	\subfigure[DDPG: 2D]{\includegraphics[width=0.49\textwidth]{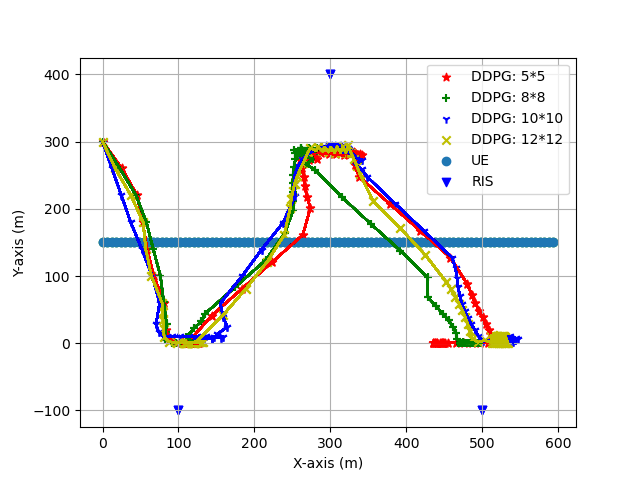}\label{ddpg_2d}}
	\caption{The trajectory of UAV obtained by DQN and DDPG-based algorithms.}
	\label{uav_tra}
\end{figure*}

\textcolor{black}{Then, we show the 3D and 2D trajectories obtained by DQN and DDPG with different number of reflecting elements in Fig.~\ref{uav_tra}. Note that in Fig.~\ref{uav_tra}, dot represents UE, and triangle represents IRS. In Fig.~\ref{dqn_3d}, it is observed that the UAV controlled by DQN starts to serve UE from the initial coordinate and finally flies to the appropriate altitude for achieving better performance. In Fig.~\ref{dqn_2d}, one can see that as the location of UE moves, the UAV flies towards to the selected IRS and remains close to it with appropriate flying actions. Also, as shown in Fig.~\ref{ddpg_3d} and Fig.\ref{ddpg_2d}, the UAV's trajectory obtained by DDPG is better than the trajectory achieved by DQN, as it always tries continuous actions.}

\begin{figure}[!htpb]
	\centering
	\includegraphics[width=3.5in]{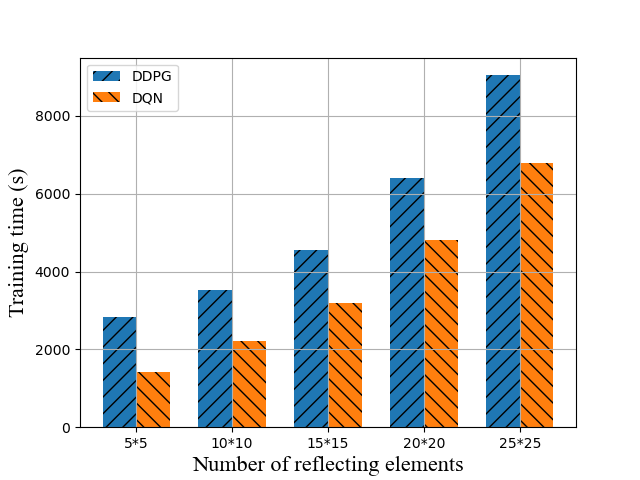}
	\caption{Training time of DQN and DDPG-based algorithms versus the number of reflecting elements of IRS.}\label{training_time}
\end{figure}

\textcolor{black}{Finally, we show the training time of DQN and DDPG-based algorithms versus the number of reflecting elements of IRS in Fig.~\ref{training_time}. Note that the training time will vary with different hardware platform. As shown in Fig.~\ref{training_time}, one can see that as the number of reflecting elements increases, the training time of DQN and DDPG increases as well. Besides, DQN consistently outperforms DDPG in terms of training time, for its simpler structure.}

\section{Conclusion}\label{conclusion}

\textcolor{black}{In this paper, we have studied the joint optimization of UAV's trajectory and passive phase shifts of reflection elements in the IRS-aided UAV communication system, with the consideration of the movement of UE and the selection of IRS. Our aim is to maximize the energy efficiency of the system, including the data rate of UE and the energy consumption of UAV. We have first proposed a DQN-based algorithm by discretizing the trajectory, which has advantage in terms of training time but has performance loss, which may be suitable for the cases that is sensitive to the training time. Then, for achieving the better performance, we have further applied a DDPG-based algorithm, which can optimize the system's variables continuously. The experimental results have proved that the proposed algorithms achieve better performance then other traditional solutions.}

\bibliographystyle{IEEEtran}
\bibliography{reference}

\begin{thebibliography}{10}
\providecommand{\url}[1]{#1}
\csname url@samestyle\endcsname
\providecommand{\newblock}{\relax}
\providecommand{\bibinfo}[2]{#2}
\providecommand{\BIBentrySTDinterwordspacing}{\spaceskip=0pt\relax}
\providecommand{\BIBentryALTinterwordstretchfactor}{4}
\providecommand{\BIBentryALTinterwordspacing}{\spaceskip=\fontdimen2\font plus
\BIBentryALTinterwordstretchfactor\fontdimen3\font minus
  \fontdimen4\font\relax}
\providecommand{\BIBforeignlanguage}[2]{{%
\expandafter\ifx\csname l@#1\endcsname\relax
\typeout{** WARNING: IEEEtran.bst: No hyphenation pattern has been}%
\typeout{** loaded for the language `#1'. Using the pattern for}%
\typeout{** the default language instead.}%
\else
\language=\csname l@#1\endcsname
\fi
#2}}
\providecommand{\BIBdecl}{\relax}
\BIBdecl

\bibitem{9275621}
F.~{Jiang}, K.~{Wang}, L.~{Dong}, C.~{Pan}, W.~{Xu}, and K.~{Yang}, ``{AI}
  driven heterogeneous {MEC} system with {UAV} assistance for dynamic
  environment: Challenges and solutions,'' \emph{IEEE Network}, vol.~35, no.~1,
  pp. 400--408, 2021.

\bibitem{9167249}
B.~{Yang}, X.~{Cao}, C.~{Yuen}, and L.~{Qian}, ``Offloading optimization in
  edge computing for deep learning enabled target tracking by
  internet-of-{UAV}s,'' \emph{IEEE Internet of Things Journal}, pp. 1--16,
  2020.

\bibitem{al2014optimal}
A.~Al-Hourani, S.~Kandeepan, and S.~Lardner, ``Optimal {LAP} altitude for
  maximum coverage,'' \emph{IEEE Wireless Communications Letters}, vol.~3,
  no.~6, pp. 569--572, 2014.

\bibitem{wu2018common}
Q.~Wu and R.~Zhang, ``Common throughput maximization in {UAV}-enabled {OFDMA}
  systems with delay consideration,'' \emph{IEEE Transactions on
  Communications}, vol.~66, no.~12, pp. 6614--6627, 2018.

\bibitem{9088229}
H.~{Ren}, C.~{Pan}, K.~{Wang}, W.~{Xu}, M.~{Elkashlan}, and A.~{Nallanathan},
  ``Joint transmit power and placement optimization for {URLLC}-enabled {UAV}
  relay systems,'' \emph{IEEE Transactions on Vehicular Technology}, pp. 1--6,
  2020.

\bibitem{zeng2019energy}
Y.~Zeng, J.~Xu, and R.~Zhang, ``Energy minimization for wireless communication
  with rotary-wing {UAV},'' \emph{IEEE Transactions on Wireless
  Communications}, vol.~18, no.~4, pp. 2329--2345, 2019.

\bibitem{8432464}
C.~H. {Liu}, Z.~{Chen}, J.~{Tang}, J.~{Xu}, and C.~{Piao}, ``Energy-efficient
  {UAV} control for effective and fair communication coverage: A deep
  reinforcement learning approach,'' \emph{IEEE Journal on Selected Areas in
  Communications}, vol.~36, no.~9, pp. 2059--2070, 2018.

\bibitem{lu2020uav}
X.~Lu, L.~Xiao, C.~Dai, and H.~Dai, ``{UAV}-aided cellular communications with
  deep reinforcement learning against jamming,'' \emph{IEEE Wireless
  Communications}, vol.~27, no.~4, pp. 48--53, 2020.

\bibitem{8764580}
Z.~{Yang}, C.~{Pan}, K.~{Wang}, and M.~{Shikh-Bahaei}, ``Energy efficient
  resource allocation in {UAV}-enabled mobile edge computing networks,''
  \emph{IEEE Transactions on Wireless Communications}, vol.~18, no.~9, pp.
  4576--4589, 2019.

\bibitem{9209079}
L.~{Wang}, K.~{Wang}, C.~{Pan}, W.~{Xu}, N.~{Aslam}, and L.~{Hanzo},
  ``Multi-agent deep reinforcement learning-based trajectory planning for
  multi-{UAV} assisted mobile edge computing,'' \emph{IEEE Transactions on
  Cognitive Communications and Networking}, vol.~7, no.~1, pp. 73--84, 2021.

\bibitem{9354996}
L.~{Wang}, K.~{Wang}, C.~{Pan}, W.~{Xu}, N.~{Aslam}, and A.~{Nallanathan},
  ``Deep reinforcement learning based dynamic trajectory control for
  {UAV}-assisted mobile edge computing,'' \emph{IEEE Transactions on Mobile
  Computing}, pp. 1--1, 2021.

\bibitem{8514812}
W.~{Huang}, Z.~{Yang}, C.~{Pan}, L.~{Pei}, M.~{Chen}, M.~{Shikh-Bahaei},
  M.~{Elkashlan}, and A.~{Nallanathan}, ``Joint power, altitude, location and
  bandwidth optimization for {UAV} with underlaid {D2D} communications,''
  \emph{IEEE Wireless Communications Letters}, vol.~8, no.~2, pp. 524--527,
  2019.

\bibitem{8119562}
C.~{Zhan}, Y.~{Zeng}, and R.~{Zhang}, ``Energy-efficient data collection in
  {UAV} enabled wireless sensor network,'' \emph{IEEE Wireless Communications
  Letters}, vol.~7, no.~3, pp. 328--331, 2018.

\bibitem{8664596}
C.~H. {Liu}, Z.~{Chen}, and Y.~{Zhan}, ``Energy-efficient distributed mobile
  crowd sensing: A deep learning approach,'' \emph{IEEE Journal on Selected
  Areas in Communications}, vol.~37, no.~6, pp. 1262--1276, 2019.

\bibitem{8365881}
J.~{Xu}, Y.~{Zeng}, and R.~{Zhang}, ``{UAV}-enabled wireless power transfer:
  Trajectory design and energy optimization,'' \emph{IEEE Transactions on
  Wireless Communications}, vol.~17, no.~8, pp. 5092--5106, 2018.

\bibitem{cui2014coding}
T.~J. Cui, M.~Q. Qi, X.~Wan, J.~Zhao, and Q.~Cheng, ``Coding metamaterials,
  digital metamaterials and programmable metamaterials,'' \emph{Light: Science
  \& Applications}, vol.~3, no.~10, p. e218, 2014.

\bibitem{di2019smart}
M.~Di~Renzo, M.~Debbah, D.-T. Phan-Huy, A.~Zappone, M.-S. Alouini, C.~Yuen,
  V.~Sciancalepore, G.~C. Alexandropoulos, J.~Hoydis, H.~Gacanin \emph{et~al.},
  ``Smart radio environments empowered by reconfigurable {AI} meta-surfaces: An
  idea whose time has come,'' \emph{EURASIP Journal on Wireless Communications
  and Networking}, vol. 2019, no.~1, pp. 1--20, 2019.

\bibitem{li2017electromagnetic}
L.~Li, T.~J. Cui, W.~Ji, S.~Liu, J.~Ding, X.~Wan, Y.~B. Li, M.~Jiang, C.-W.
  Qiu, and S.~Zhang, ``Electromagnetic reprogrammable coding-metasurface
  holograms,'' \emph{Nature communications}, vol.~8, no.~1, pp. 1--7, 2017.

\bibitem{9140329}
M.~{Di Renzo}, A.~{Zappone}, M.~{Debbah}, M.~S. {Alouini}, C.~{Yuen}, J.~{de
  Rosny}, and S.~{Tretyakov}, ``Smart radio environments empowered by
  reconfigurable intelligent surfaces: How it works, state of research, and the
  road ahead,'' \emph{IEEE Journal on Selected Areas in Communications},
  vol.~38, no.~11, pp. 2450--2525, 2020.

\bibitem{wu2019intelligent}
Q.~Wu and R.~Zhang, ``Intelligent reflecting surface enhanced wireless network
  via joint active and passive beamforming,'' \emph{IEEE Transactions on
  Wireless Communications}, vol.~18, no.~11, pp. 5394--5409, 2019.

\bibitem{6942282}
K.~{Wang}, Y.~{Chen}, and M.~{Di Renzo}, ``Outage probability of dual-hop
  selective {AF} with randomly distributed and fixed interferers,'' \emph{IEEE
  Transactions on Vehicular Technology}, vol.~64, no.~10, pp. 4603--4616, 2015.

\bibitem{huang2020holographic}
C.~Huang, S.~Hu, G.~C. Alexandropoulos, A.~Zappone, C.~Yuen, R.~Zhang,
  M.~Di~Renzo, and M.~Debbah, ``Holographic {MIMO} surfaces for {6G} wireless
  networks: Opportunities, challenges, and trends,'' \emph{IEEE Wireless
  Communications}, vol.~27, no.~5, pp. 118--125, 2020.

\bibitem{8647620}
Q.~{Wu} and R.~{Zhang}, ``Intelligent reflecting surface enhanced wireless
  network: Joint active and passive beamforming design,'' in \emph{2018 IEEE
  Global Communications Conference (GLOBECOM)}, 2018, pp. 1--6.

\bibitem{yang2020intelligent}
Y.~Yang, B.~Zheng, S.~Zhang, and R.~Zhang, ``Intelligent reflecting surface
  meets {OFDM}: Protocol design and rate maximization,'' \emph{IEEE
  Transactions on Communications}, 2020.

\bibitem{9014322}
X.~Yu, D.~Xu, and R.~Schober, ``Enabling secure wireless communications via
  intelligent reflecting surfaces,'' in \emph{2019 IEEE Global Communications
  Conference (GLOBECOM)}, 2019, pp. 1--6.

\bibitem{zhou2020intelligent}
G.~Zhou, C.~Pan, H.~Ren, K.~Wang, and A.~Nallanathan, ``Intelligent reflecting
  surface aided multigroup multicast {MISO} communication systems,'' \emph{IEEE
  Transactions on Signal Processing}, 2020.

\bibitem{pan2020multicell}
C.~Pan, H.~Ren, K.~Wang, W.~Xu, M.~Elkashlan, L.~Hanzo, and A.~Nallanathan,
  ``Multicell {MIMO} communications relying on intelligent reflecting
  surfaces,'' \emph{IEEE Transactions on Wireless Communications}, 2020.

\bibitem{9110849}
C.~Pan, H.~Ren, K.~Wang, M.~Elkashlan, A.~Nallanathan, J.~Wang, and L.~Hanzo,
  ``Intelligent reflecting surface aided {MIMO} broadcasting for simultaneous
  wireless information and power transfer,'' \emph{IEEE Journal on Selected
  Areas in Communications}, vol.~38, no.~8, pp. 1719--1734, 2020.

\bibitem{9133107}
T.~{Bai}, C.~{Pan}, Y.~{Deng}, M.~{Elkashlan}, A.~{Nallanathan}, and
  L.~{Hanzo}, ``Latency minimization for intelligent reflecting surface aided
  mobile edge computing,'' \emph{IEEE Journal on Selected Areas in
  Communications}, pp. 1--17, 2020.

\bibitem{9148961}
S.~{Abeywickrama}, R.~{Zhang}, and C.~{Yuen}, ``Intelligent reflecting surface:
  Practical phase shift model and beamforming optimization,'' in \emph{ICC 2020
  - 2020 IEEE International Conference on Communications (ICC)}, 2020, pp.
  1--6.

\bibitem{zappone2020overhead}
A.~Zappone, M.~Di~Renzo, F.~Shams, X.~Qian, and M.~Debbah, ``Overhead-aware
  design of reconfigurable intelligent surfaces in smart radio environments,''
  \emph{arXiv preprint arXiv:2003.02538}, 2020.

\bibitem{9110869}
C.~Huang, R.~Mo, and C.~Yuen, ``Reconfigurable intelligent surface assisted
  multiuser miso systems exploiting deep reinforcement learning,'' \emph{IEEE
  Journal on Selected Areas in Communications}, vol.~38, no.~8, pp. 1839--1850,
  2020.

\bibitem{li2020reconfigurable}
S.~Li, B.~Duo, X.~Yuan, Y.-C. Liang, and M.~Di~Renzo, ``Reconfigurable
  intelligent surface assisted {UAV} communication: Joint trajectory design and
  passive beamforming,'' \emph{IEEE Wireless Communications Letters}, vol.~9,
  no.~5, pp. 716--720, 2020.

\bibitem{ma2019enhancing}
D.~Ma, M.~Ding, and M.~Hassan, ``Enhancing cellular communications for {UAV}s
  via intelligent reflective surface,'' \emph{arXiv preprint arXiv:1911.07631},
  2019.

\bibitem{10.5555/3312046}
R.~S. Sutton and A.~G. Barto, \emph{Reinforcement Learning: An
  Introduction}.\hskip 1em plus 0.5em minus 0.4em\relax Cambridge, MA, USA: A
  Bradford Book, 2018.

\bibitem{mnih2015human}
V.~Mnih, K.~Kavukcuoglu, D.~Silver, A.~A. Rusu, J.~Veness, M.~G. Bellemare,
  A.~Graves, M.~Riedmiller, A.~K. Fidjeland, G.~Ostrovski \emph{et~al.},
  ``Human-level control through deep reinforcement learning,'' \emph{Nature},
  vol. 518, no. 7540, pp. 529--533, 2015.

\bibitem{lillicrap2015continuous}
T.~P. Lillicrap, J.~J. Hunt, A.~Pritzel, N.~Heess, T.~Erez, Y.~Tassa,
  D.~Silver, and D.~Wierstra, ``Continuous control with deep reinforcement
  learning,'' \emph{arXiv preprint arXiv:1509.02971}, 2015.

\bibitem{6313077}
A.~G. {Barto}, R.~S. {Sutton}, and C.~W. {Anderson}, ``Neuronlike adaptive
  elements that can solve difficult learning control problems,'' \emph{IEEE
  Transactions on Systems, Man, and Cybernetics}, vol. SMC-13, no.~5, pp.
  834--846, 1983.

\bibitem{ren2019achievable}
H.~Ren, C.~Pan, K.~Wang, Y.~Deng, M.~Elkashlan, and A.~Nallanathan,
  ``Achievable data rate for urllc-enabled {UAV} systems with {3-D} channel
  model,'' \emph{IEEE Wireless Communications Letters}, vol.~8, no.~6, pp.
  1587--1590, 2019.

\bibitem{wei2020sum}
Z.~Wei, Y.~Cai, Z.~Sun, D.~W.~K. Ng, J.~Yuan, M.~Zhou, and L.~Sun, ``Sum-rate
  maximization for {IRS}-assisted {UAV} {OFDMA} communication systems,''
  \emph{IEEE Transactions on Wireless Communications}, vol.~20, no.~4, pp.
  2530--2550, 2020.

\end{thebibliography}

\end{document}